%% file: main.tex
\documentclass[runningheads]{article}
\usepackage[section]{placeins}
\usepackage{graphicx}
\usepackage{authblk}
\providecommand{\keywords}[1]{\textbf{\textit{Keywords---}} #1}
\begin{document}
\date{}
\title{Exploring Scientific Application Performance Using Large Scale Object Storage}
%

%

%

\author[1]{Steven Wei-der Chien}
\author[1]{Stefano Markidis}
\author[1]{Rami Karim}
\author[1]{\\Erwin Laure}
\author[2]{Sai Narasimhamurthy}
\affil[1]{KTH Royal Institute of Technology, Sweden}
\affil[2]{Seagate Systems UK, UK}

\maketitle              
\begin{abstract}
One of the major performance and scalability bottlenecks in large scientific applications is parallel reading and writing to supercomputer I/O systems. The usage of parallel file systems and consistency requirements of POSIX, that all the traditional HPC parallel I/O interfaces adhere to, pose limitations to the scalability of scientific applications. Object storage is a widely used storage technology in cloud computing and is more frequently proposed for HPC workload to address and improve the current scalability and performance of I/O in scientific applications. While object storage is a promising technology, it is still unclear how scientific applications will use object storage and what the main performance benefits will be. This work addresses these questions, by emulating an object storage used by a traditional scientific application and evaluating potential performance benefits. We show that scientific applications can benefit from the usage of object storage on large scales.
\end{abstract}
\keywords{Scientific Applications,  Object Storage, Parallel I/O, HPC, HDF5}

\input{introduction}

\input{related_work}
\input{method}

\input{results}
\input{conclusion}

\section*{Acknowledgments}
Funding for the work is received from the European Commission H2020 program, Grant Agreement No. 671500 (SAGE).
%
%
%
%

\bibliographystyle{splncs04}
\bibliography{main}
\end{document}

%% file: introduction.tex
\section{Introduction}
Parallel I/O is becoming one of the most serious performance bottlenecks in HPC applications as the number of processes writing/reading to/from the supercomputer I/O system keeps increasing at a considerable pace. An exascale supercomputer will likely support billions of processes~\cite{bergman2008exascale} that can potentially access and update shared files, all at the same time. The implementation of existing HPC parallel interfaces, such as MPI I/O, HDF5 and NetCDF are all based on and complaint to the POSIX standard. The POSIX standard requires strong consistency when accessing and updating a file. In a parallel environment, strong consistency is achieved by a process acquiring a lock on the file, completing the operation and releasing the lock. One reason for the performance bottleneck of HPC parallel I/O interfaces is the strong consistency POSIX requirement, and its implementation. 

One of the possible disruptive solutions to address the lack of scalability of traditional parallel I/O would be the adoption of object storage technology~\cite{factor2005object}. Object storage is a well-spread technology in cloud computing and is currently being utilized by large tech companies. For instance, Amazon and Google, to mention a few, have implemented Amazon S3 and Google Cloud object storage; services which are used by many companies today. Object storage abandons traditional POSIX I/O concepts, such as directories, files, and certain file operations. Unlike many other parallel file systems, object storage provides a single flat global name space and supports only a few operations, among which are the \textsf{PUT} and \textsf{GET} operations. Performance scalability of object storage is partly due to the concept of \emph{object immutability} and also due to the semantics of the \textsf{PUT} and \textsf{GET} operations. Objects are \emph{immutable}, as in-place changes to the data in the object are not possible. A \textsf{PUT} operation creates an object, adds data to it and returns an object Universally Unique Identifier (UUID). A simple hash function or a combination of them can then be used to determine the location of the object in Object Storage Device (OSD). The main two advantages of object storage when compared to traditional approaches are:
\begin{enumerate}
\item Because objects are immutable, it is impossible for a node to write to an object that is being read by others. This allows removal of locks while reading data from an object, providing a lock-free data access.
\item Because physical object locations can be determined by object UUID and hashing function, it possible to directly access data without any locational metadata.
\end{enumerate}
The main limitation of the object storage is that it requires additional software for metadata. In fact, metadata, such as the object name, creation time, etc., that is not comprised in the object storage needs to be stored and managed outside the object storage. For this reason, objects stores are usually equipped with a key-value store, providing users with a front-end interface and mapping object UUID to metadata.

\begin{figure}
  \begin{center} 
  \includegraphics[width=0.9\linewidth]{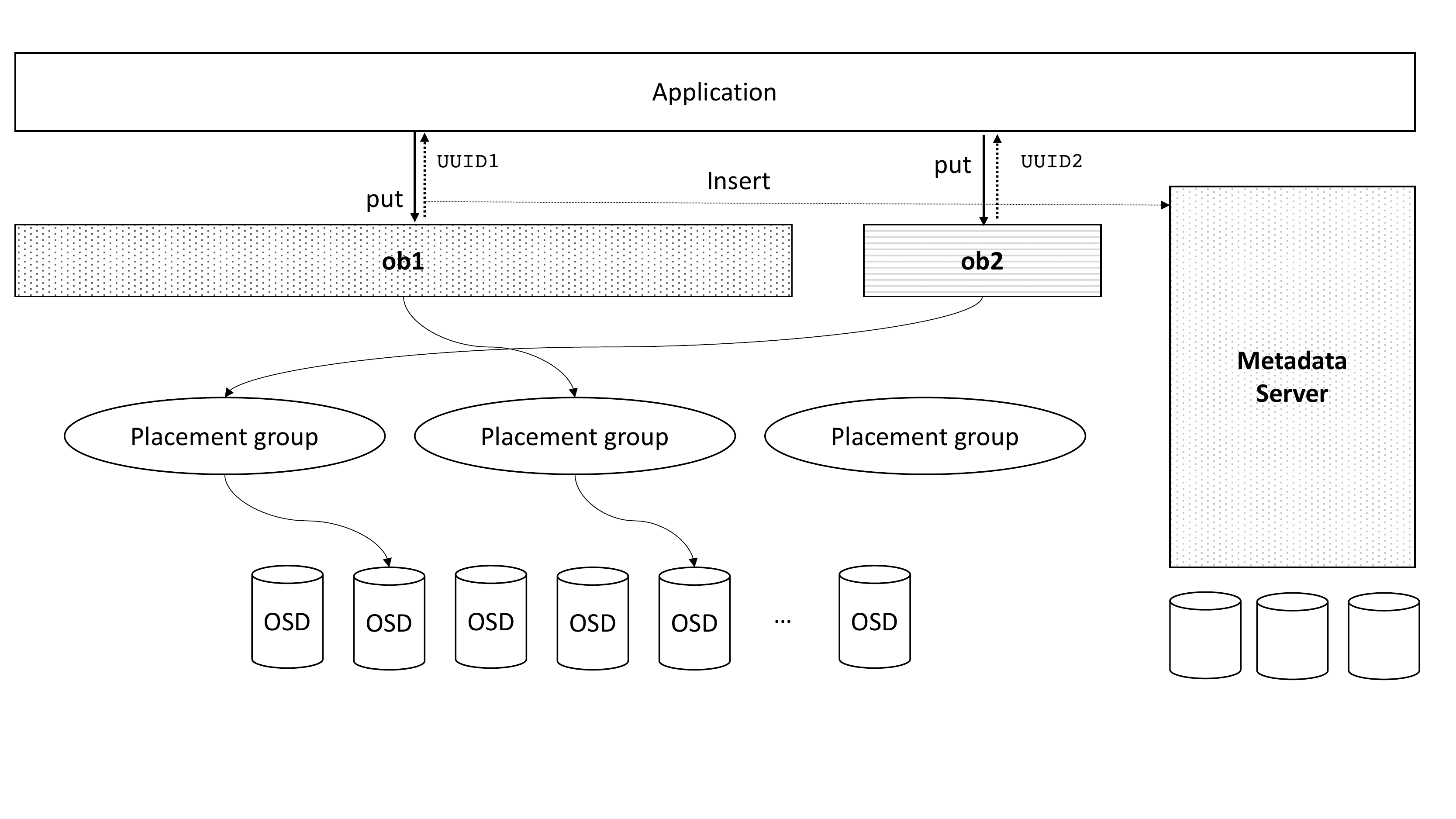}
  \caption{Simplified diagram of an application writing two objects to the object store, similar to CEPH~\cite{weil2006ceph}.}
 \label{fig:Figure1}
 \end{center} 
\end{figure}

A simplified diagram of object storage is presented in Fig.~\ref{fig:Figure1}, showing an application putting two objects in the object storage, similar to CEPH object storage~\cite{weil2006ceph}. With the \textsf{PUT} operation, two objects are created and their UUIDs are retrieved. A hash function is used to determine the placement group (an object pool) and then the physical location on the OSDs. The \textsf{PUT} operation also inserts associated metadata to the metadata server.

While object storage is a promising technology that could potentially replace parallel file systems in the future, it is unclear how current scientific applications running on supercomputers will use object storage and what the potential benefits are. To the best of our knowledge, there are no large scale supercomputers that are directly using object storage. For this reason, our goal is to take an application with similar workload to typical HPC applications, and emulate object storage on it. We emulate object storage to write both individual and shared objects and to compare the I/O performance with existing parallel HDF5 implementations~\cite{folk2011overview}.

The paper makes the following contributions:
\begin{itemize}
\item We propose a methodology to evaluate the scalability of object storage at scale and develop a simple emulator to write and read objects on an object store serving large supercomputers.
\item We deploy the object storage emulator in a representative massively parallel application and measure the I/O performance.
\item We analyze the emulated object storage performance at scale, compare it with performance of parallel HDF5 and evaluate the object storage potential for extreme scale systems.
\end{itemize}

The paper is organized as follows. First, we provide background to this study and present related work in Section~\ref{sec:relwork}. Section~\ref{sec:methods} presents the design and implementation of an emulator for object storage at scale. Section \ref{sec:experiment} introduces the benchmarking environment. Section~\ref{sec:results} presents the results. Finally, Section~\ref{sec:conclusions} summarizes the results, discusses the limitations of this work and outlines future work.

%% file: related_work.tex
\section{Background \& Related Work}
\label{sec:relwork}

In this section, we provide an overview of common parallel I/O libraries and file systems, together with the related work. 

{\bf POSIX I/O.} POSIX (Portable Operating System Interface) is a specification defined by IEEE Computer Society for a standardized operating system interface and environment. Among other operations and concepts, POSIX defines the interaction between file descriptors and standard I/O streams~\cite{posix}. From a programmer's perspective, POSIX I/O is characterized by the following interfaces: \textsf{open()}, \textsf{close()}, \textsf{read()}, \textsf{write()}, and \textsf{lseek()}. Additionally, the POSIX standard specifies the semantics of such operations. POSIX I/O is stateful and requires the operating system to maintain persistent states. In order to modify a file one must first \textsf{open} a descriptor, \textsf{seek} a location and \textsf{read} or \textsf{write} from there. File descriptors are not shared between processes and the system must maintain every descriptor opened by all processes.

Another feature of POSIX I/O is the strict consistency requirement. POSIX defines that after a successful \textsf{write()} call, any subsequent \textsf{read()} operations from the byte positions in the modified file must return data written by that \textsf{write()} operation~\cite{posix}. The consistency semantics is often implemented with some form of locking mechanism~\cite{GPFS}. In the case of parallel file systems, this is often implemented via distributed locking~\cite{distributed-token}.

{\bf Parallel File Systems.} Parallel file systems such as \emph{Lustre}~\cite{lustre}, \emph{GPFS}~\cite{GPFS} and \emph{VPFS}~\cite{pvfs} are created to support parallel I/O in a cluster environment. These file systems implement I/O forwarding at an extra layer between the storage system and computing system, which handles I/O on behalf of the computing systems. One of the widely adopted parallel file systems is Lustre. Lustre Metadata Servers (MDS) handles information such as physical locations, file names, permissions and timestamps. Data is striped and sent to different Object Stores Servers (OSS)~\cite{schwan2003lustre}. Although Lustre is object store based, it exposes itself through POSIX I/O interface and is near POSIX compliant. POSIX consistency semantics is enforced through distributed locking. One performance bottleneck of Lustre lies on metadata management. Another bottleneck lies in file locking, which it is required to preserve consistency. Excessive striping can also negatively impact performance~\cite{lioprof}\cite{lustre-file-joining}. Performance of parallel writing to a single file can be improved by explicit configuration of striping, yet it increases failure risk and worsen performance of writes to non-shared files~\cite{MPI-IO-performance}.

{\bf Parallel I/O Libraries.}  It is common to use parallel file systems with parallel I/O libraries. \emph{MPI-IO}~\cite{MPI-IO}, \emph{Parallel HDF5}~\cite{hdf5} and \emph{Parallel netCDF}~\cite{parallel-netcdf} are parallel I/O libraries that work on top of parallel file systems. MPI-IO aims to provide a portable interface which addresses common parallel I/O patterns, such as collective I/O and non-contiguous access. MPI-IO exposes several POSIX like I/O interfaces with relaxed consistency requirements. MPI-IO guarantees that a write from one process is immediately visible to processes in the same communicator group in which the file was opened with atomic mode. Otherwise the content is only visible after explicit synchronization. Many I/O libraries, such as HDF5 and Parallel netCDF, are built upon MPI-IO to take advantage of portable parallel I/O. However, MPI-IO provides little performance improvement for contiguous access~\cite{MPI-IO-performance}. Due to its similarity to POSIX I/O, the performance of these two APIs is often similar.

{\bf Object Storage System.} Object storage system is an architecture which manages data as objects instead of files~\cite{factor2005object}. Due to their scalability, object stores are widely adopted in cloud based systems. Object store operations are stateless and in object store semantics there are only two basic operations: \textsf{GET} and \textsf{PUT}. A \textsf{PUT} operation returns an ID which uniquely represents the object. Object store implementations usually provides a facility to map an assigned name to an ID, together with metadata which describes the object. This is often implemented with a key-value store. All objects are stored without structure and clients communicate directly to the storage node where data physically resides without requiring location lookup by hashing the object's ID~\cite{weil2006ceph}. Objects are immutable and it is impossible to concurrently create or update the same object. This eliminates the bottleneck caused by locking. In contrary to POSIX I/O, object stores support a weak form of consistency: eventual consistency. This means that a successfully returned \textsf{PUT} operation does not necessary require that the object will be visible immediately. Deterministic placement of object through ID hashing leads to the elimination bottleneck due to lookup.

\emph{CEPH}~\cite{weil2006ceph} is one of the most commonly known object storage system. It exposes itself as through a POSIX interface and at the same time provides a number of POSIX I/O extensions which provides relaxed consistency. Unlike Lustre, any party can compute the physical location of an object by hashing its ID. For this reason, location metadata is completely eliminated. This reduces the stress on the metadata cluster. Additionally, it is possible to manipulate the underlying object store directly through librados~\cite{librados}. Additional emerging object storages, targeting HPC workloads, are Seagate's Mero~\cite{narasimhamurthy2018sage}\cite{SageComputingFront}, DAOS~\cite{breitenfeld2017daos} and DDN's Web Object Store (WOS)~\cite{wos}. Studies have also been made on how HPC applications can interact with these transaction based storage systems~\cite{rivas2017mpi}. The adoption of these systems enabled a wider range of underlying storage technologies to be used, such as hybrid-memory storage systems and Non-volatile memory storage systems~\cite{lockwood2017storage}\cite{peng2017exploring}\cite{peng2016exploring}.

{\bf I/O Pattern in Scientific Applications.} The majority of scientific applications perform a large number of write operations to preserve intermediate states and final outcome of simulation variables for post-processing (visualization and data analysis) and check-pointing. In scientific applications, these operations occur typically at a given computational cycle, defined by users. Because of the computational cost of parallel I/O, I/O operations are kept at minimum in scientific applications. Typically, these outputs are only for the purpose of archiving, later post-processing and check-pointing for restarting simulations.

Parallel I/O operations either:
\begin{enumerate}
\item Write/Read one file per process (independent parallel I/O), or
\item Write/Read to the same shared file (cooperative parallel I/O). In this case, parallel I/O is performed using parallel I/O libraries, such as MPI-IO, Parallel HDF5 and NETCDF.
\end{enumerate}
 
Studies have shown that parallel write to the same file often results in worse performance than writing to individual non-shared files~\cite{MPI-IO-performance}\cite{byna2017tuning}\cite{workload-pattern}, implying that existing parallel file systems are not well suited for parallel I/O. For Lustre, custom striping configuration can result in similar performance between the two approaches, but also results in higher failure risk.

In terms of concurrent write, it is rarely the case that a process requires the latest update through a read operation immediately after a write operation. This can be efficiently implemented through MPI point-to-point or collective where data is in-memory if data sharing is needed. Therefore, the strict read-after-write consistency requirement imposed by POSIX I/O is rarely required.

By decoupling metadata management and relaxing consistency requirements, object stores can potentially provide extreme scalability for parallel I/O. Scalability is achieved through deterministic placements and lock-free accesses. Since most scientific applications are write intensive and do no rely on POSIX consistency guarantees, we argue that object stores will be extremely valuable for scientific applications.

%% file: method.tex
\section{Emulating Scientific Applications Using Object Storage}
\label{sec:methods}

\begin{figure}
	\begin{center} 
		\includegraphics[width=0.9\linewidth]{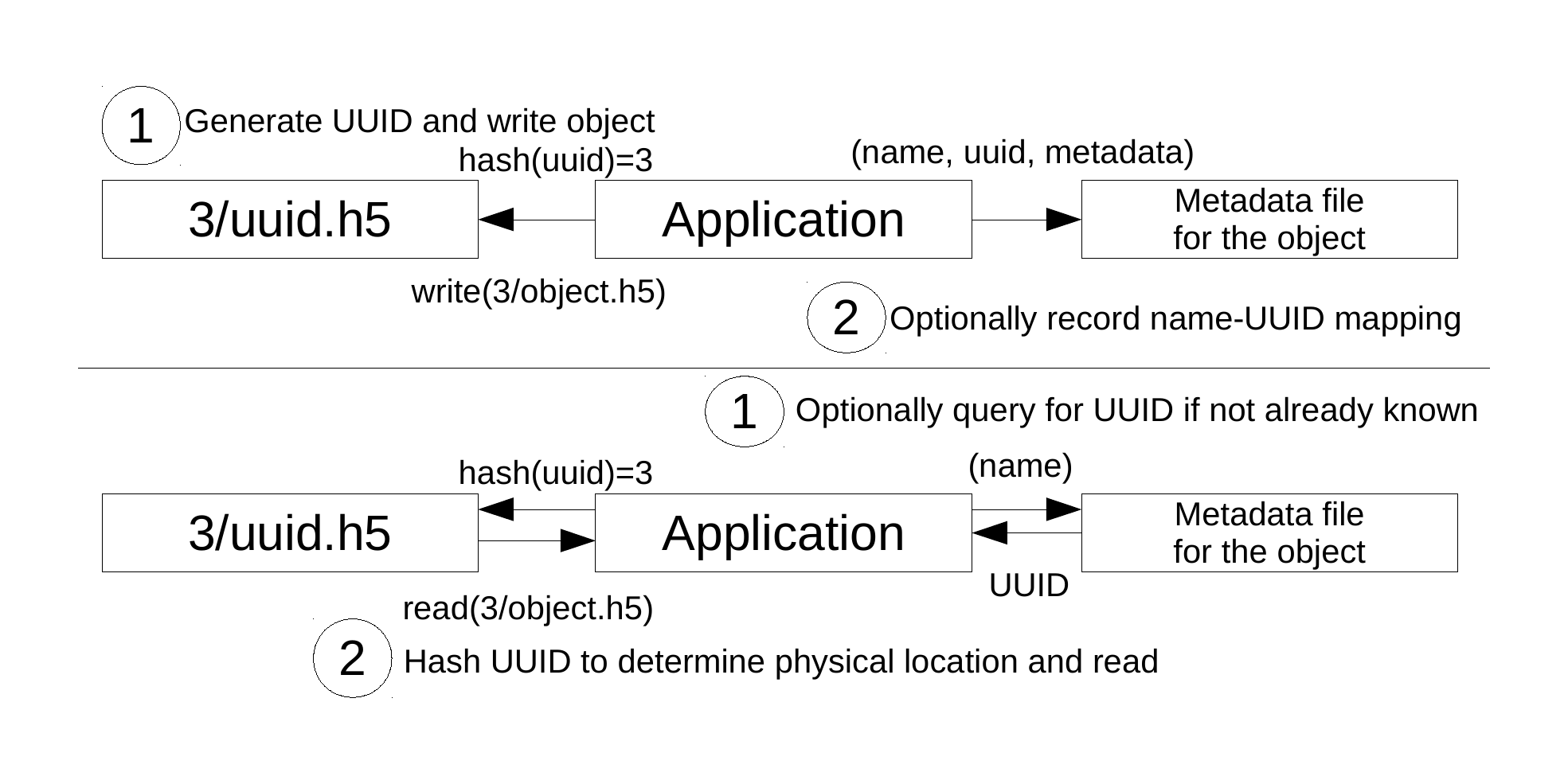}
		\caption{Illustration of how our emulator mimics object store \textsf{PUT} (top) and \textsf{GET} (bottom) operations in an application.}
		\label{fig:Figure5}
	\end{center} 
\end{figure}

Our goal is to assess the impact of object storage system on supercomputers in HPC scientific applications. In particular, we are interested in how object stores can improve scalability of such applications. Yet, HPC-ready object stores are not widely adopted. For this reason, we have designed and implemented a simple library that emulates the workings of an object storage system. Specifically, we emulate four key features of object storage systems, namely: flat namespace structure, object immutability, deterministic object placement and metadata management. We implement the API according to object store semantics. A \textsf{GET} operation retrieves an object and a \textsf{PUT} operation creates a new object as shown in Fig~\ref{fig:Figure5}. Metadata are stored as serialized binary files to mimic a key-value store where the file name is the key. Furthermore, we support chunking operations. This implies that it is possible to create an object concurrently by different processes.

\subsection{Emulator Implementation}

\begin{figure}
				\begin{center} 
					\includegraphics[width=0.9\linewidth]{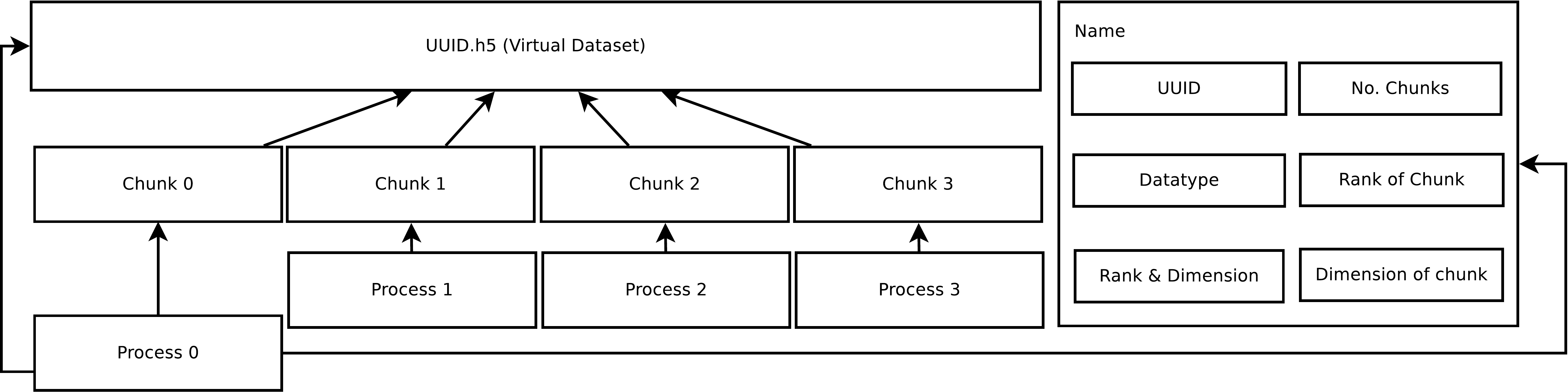}
					\caption{Illustration of how the emulator uses object chunks and metadata.}
					\label{fig:Figure3}
				\end{center} 
			\end{figure}

We implement an emulator to mimic an object storage serving a large scale supercomputer as a C library. For the purpose of the experiment, we support the storage of multidimensional arrays in 64-bit double, 32-bit float and 32-bit integer data-type as objects. Objects are stored in one or more HDF5 files. These HDF5 files are represented by UUID and a part number. Metadata are represented in protobuf serialized data format. Furthermore, we emulate different Object Storage Devices (OSD) as different folders. Our emulator employs a weak form of consistency: eventual consistency. This means that an object being written will not be immediately visible to other processes, but will eventually be. We define that an object is visible after the metadata is written and synchronized to disk. The writing of metadata is performed after all data is successfully written. This is to support wait-free read by other processes to an object with the same name. In this way, processes will retrieve the ID of the object from existing metadata, which point to an existing object, while the new object is being written to another location with another ID. We also write a HDF5 virtual dataset which links all the chunks together and present a unified view of the entire object. HDF5 virtual datasets can be opened as a single dataset with existing HDF5 dataset APIs as if it is one single file. How our emulator writes an object chunk and correspondent metadata is shown in Fig.~\ref{fig:Figure3}.

\subsubsection{Objects Creation}

A new object is created with a \textsf{PUT} operation. A \textsf{PUT} operation receives object name, data and metadata which describes the object, and performs operations. We support multidimensional data in a variety of datatypes. Additionally, we represent descriptive information such as size and dimension as metadata. When a \textsf{PUT} request is received, a new UUID is generated to uniquely represent the object. We use a hash function to determine in which object storage device the object will the be placed. A new HDF5 dataset will be created to store the structured input data and stored in an HDF5 file with the UUID as filename.

\subsubsection{Object chunking}

\begin{figure}
		\centering
	\includegraphics[width=0.9\linewidth]{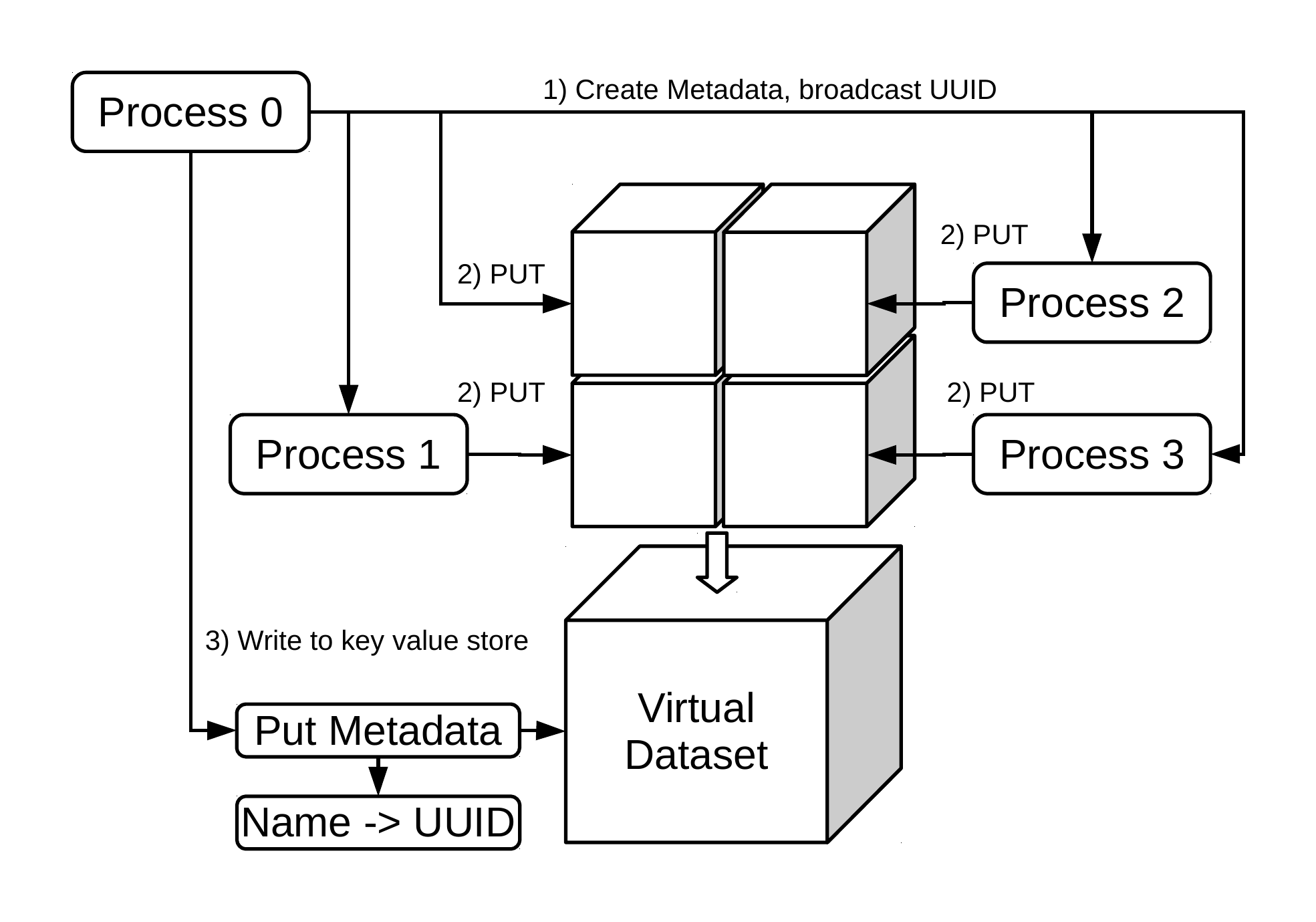}
	\caption{In our emulated object storage, object chunks are stored as HDF5 files and metadata as Virtual Data Sets (VDS).}
	\label{fig:vds-chunks-workflow}
\end{figure}

During object chunking, the object is divided into equal sized portions and stored individually with many HDF5 files. Multidimensional chunks are supported. A chunk ID is appended to the filename for reconstruction. Immediately before the metadata is being written, a HDF5 Virtual Dataset (VDS) is created to provide a high level overview of the object being written~\cite{byna2017tuning}\cite{hdf5-vds}. Thus, the object chunks can be retrieved as a single object thereafter. Fig.~\ref{fig:vds-chunks-workflow} shows our object store design, where individual object chunks are represented as individual HDF5 files and are linked via VDS according to part number.

\subsubsection{Metadata Management}

We represent object metadata using \textsf{protobuf}, which is a serialization mechanism developed by Google. Our emulated object storage supports data in multi-dimensional tensors. So, in our particular case, we store the rank of the tensor, the size for each tensor dimension and the UUID. If the object is chunked, then the chunk size, dimension and chunk count will be also stored. The protobuf object will be serialized and written to a temporary file on disk. After the file is synchronized to disk, we perform a POSIX \textsf{rename()} and synchronization to rename the temporary file to the user defined object name. This ensures that a third party client who is accessing the metadata file with the same name will either get a new or an old copy of the metadata. A client who is holding a file descriptor to the old metadata will still be able to read the old copy of the metadata. In the case where multiple clients are creating different chunks of the same object, the master process must initiate a \textsf{PUT} process by obtaining a copy of the metadata and UUID from the library. The UUID will be broadcast to other processes and they can perform their own chunked \textsf{PUT} by supplying the data, UUID and part number. When all processes have completed their respective chunked \textsf{PUT}, the master process performs a \textsf{commit} to write the metadata to disk so that the new object will be visible. It is the application's responsibility to ensure that all chunked \textsf{PUT} operations by different processes are complete. When two processes perform \textsf{PUT}s with the same object name concurrently, the most up-to-date version of the object is the object of the last process writing the metadata.

%% file: results.tex
\section{Experimental Environment}
\label{sec:experiment} 

Our experiments are performed on the Beskow supercomputer at KTH. Beskow is a Cray XC40 system, consisting of 2,060 compute nodes, equipped with two Xeon E5-2698v3 Haswell 2.3 GHz CPUs (16 cores per CPU) per node and high speed network Cray Aries. The storage employs a Lustre parallel file system (client v2.5.2) with 165 OST servers. Beskow OS is SUSE LINUX (Release 11).We use GCC version 4.9.1, Cray-MPICH v7.0.4 and HDF5 v1.10.1. 

To measure the I/O performance, we use the Darshan profiler~\cite{darshan}. Darshan is a low-overhead tool to investigate the I/O performance of parallel applications. Darshan provides bandwidth and measured time spent on I/O. Measurement is done at MPI I/O and POSIX level. For this reason, the Darshan tool is capable of profiling both our emulated library and parallel HDF5 as our library is implemented through HDF5 which is based on POSIX I/O; and parallel HDF5 which is based on MPI I/O.

\begin{figure}[h]
	\begin{center} 
		\includegraphics[width=0.9\linewidth]{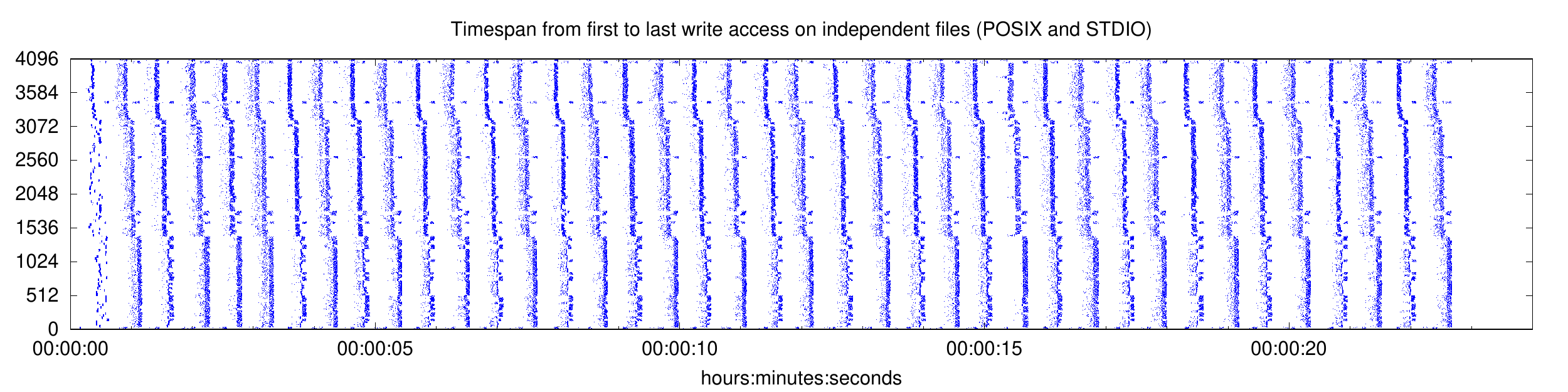}
		\caption{Different processes writing to individual object chunks.}
		\label{fig:individual-io-pattern-figure}
	\end{center} 
\end{figure}

We implemented a skeleton application iterating over 200 computational cycles. Every five cycles, we perform parallel I/O. Therefore, the whole execution cycle consists of 40 I/O phases. The computation and I/O phases take approximately 75\% and 25\% of the total execution time respectively. We perform weak scaling test, keeping the amount of data to be written by each process constant while varying number of processes. For every I/O phase, we output a two dimensional integer array where the size is a multiple of chunk size with the number of processes. A broadcast of UUID is performed by rank 0, processes write their chunk and sign-in to a barrier. Rank 0 commits metadata to disk after all processes are signed-in. To compare the I/O performance with parallel HDF5, a separate experiment that utilizes HDF5 was created. The set-up consists of the same skeleton application performing calculations at each iteration and I/O with parallel HDF5 to write to a shared file every five cycles. MPI hints are provided to utilize Lustre striping with a stripe count equal to the number of MPI processes in use divided by the number of processes per stripe.

We created test configuration with different chunk sizes. We tested $16 \times 4096$, $32 \times 4096$ and $64 \times 4096$. For parallel HDF5, we additionally test for different number of processes per stripe. We tested 16, 32 and 64, where 32 is the number of processors on one computing node of Beskow. For each configuration we repeat the tests 5 times and report the median, minimum and maximum bandwidth in MiB/s. For time spent on I/O operation we report the average value in seconds.

\begin{figure}[h]
	\begin{center} 
		\includegraphics[width=1\linewidth]{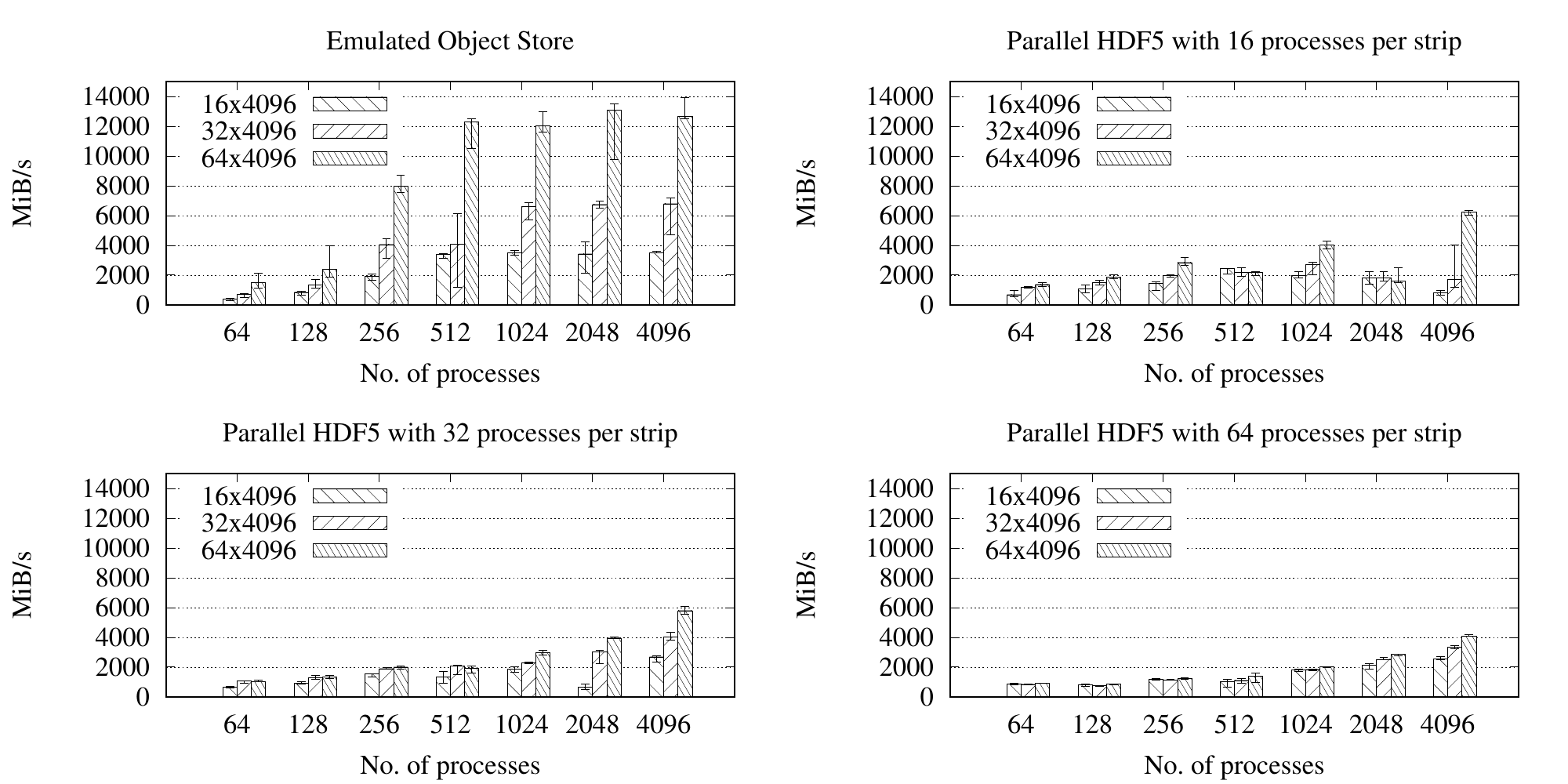}
		\caption{Median bandwidth in MiB/s measured by Darshan for different configurations. Maximum and minimum bandwidth recorded are represented by error bars. Scaling of workload results in increase of bandwidth in both methods but on a different scale.}
		\label{fig:all-bandwidth}
	\end{center}
\end{figure}

\section{Evaluation}
\label{sec:results}

\begin{figure}[h]
	\begin{center} 
		\includegraphics[width=0.9\linewidth]{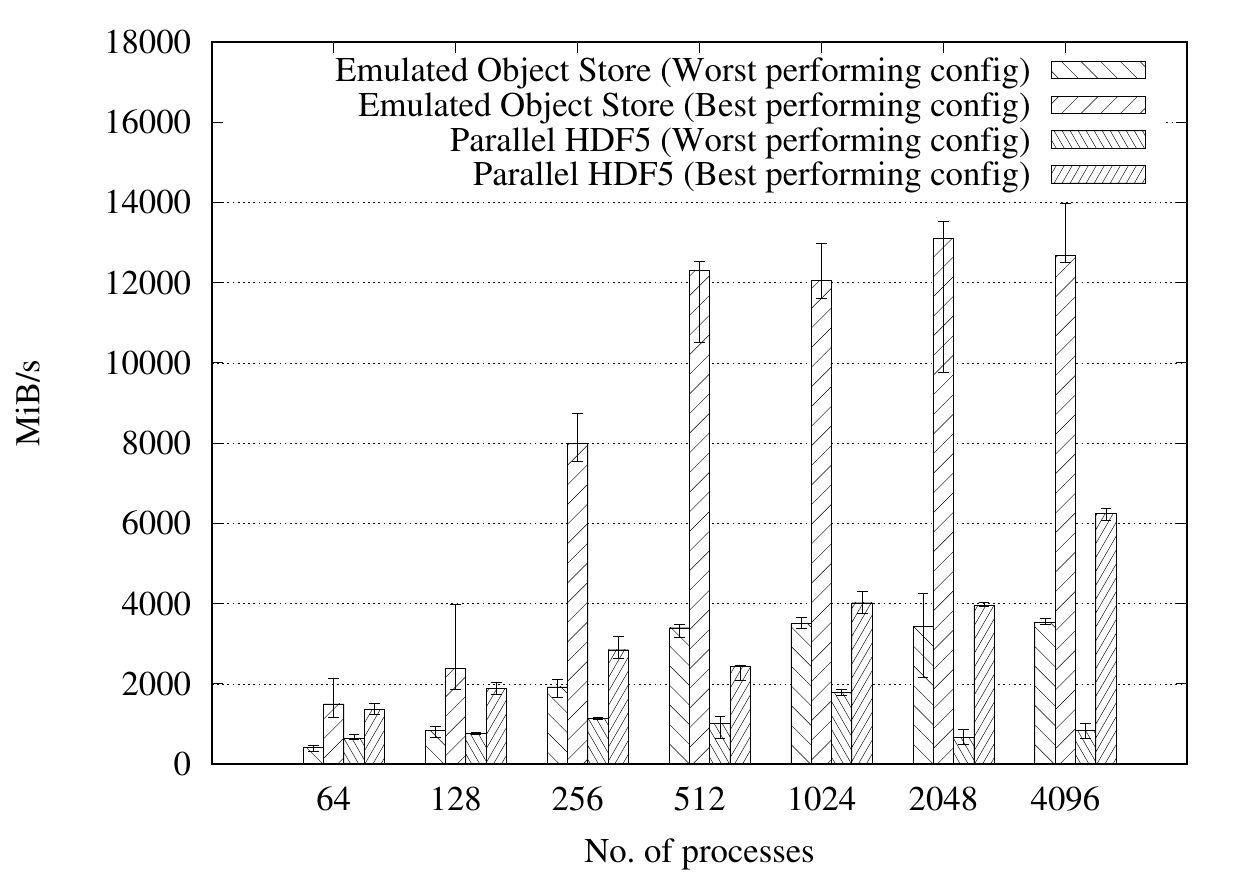}
		\caption{Configurations with maximal and minimal bandwidth in MiB/s measured by Darshan among different configurations. Both methods provide comparable bandwidth with small number of processes.}
		\label{fig:bandwidth-figure}
	\end{center} 
\end{figure}

We perform scaling tests up to 4,096 processes. We measure both time spent on I/O operations and bandwidth of write operations. Fig.~\ref{fig:individual-io-pattern-figure} shows I/O patterns of different processes of a particular configuration using our emulator where processes perform putting of their data chunk. Fig.~\ref{fig:all-bandwidth} shows the bandwidth under different chunk sizes and stripe configuration by the application with the two I/O methods: emulated object storage and parallel HDF5 to a shared file. Our emulator outperforms parallel HDF5 in terms of bandwidth. Comparing to the emulator, parallel HDF5 only provides moderate scaling in bandwidth. We also noticed that the performance of parallel HDF5 is extremely sensitive to configurations. Fig.~\ref{fig:bandwidth-figure} shows the maximal and minimal bandwidth measured from writing operations among all configurations.

We find that our emulated object store implementation provides better scalability than writing to shared file. Although still outperforming parallel HDF5, after more than 2,048 processes are used, we observe saturation and slight decline in bandwidth. We also observed that the bandwidth scales together with increasing chunk size with our emulator. The same is observed with parallel HDF5. Yet the scaling comparing with the emulator is only moderate.
\begin{figure}[h]
	\begin{center} 
		\includegraphics[width=0.8\linewidth]{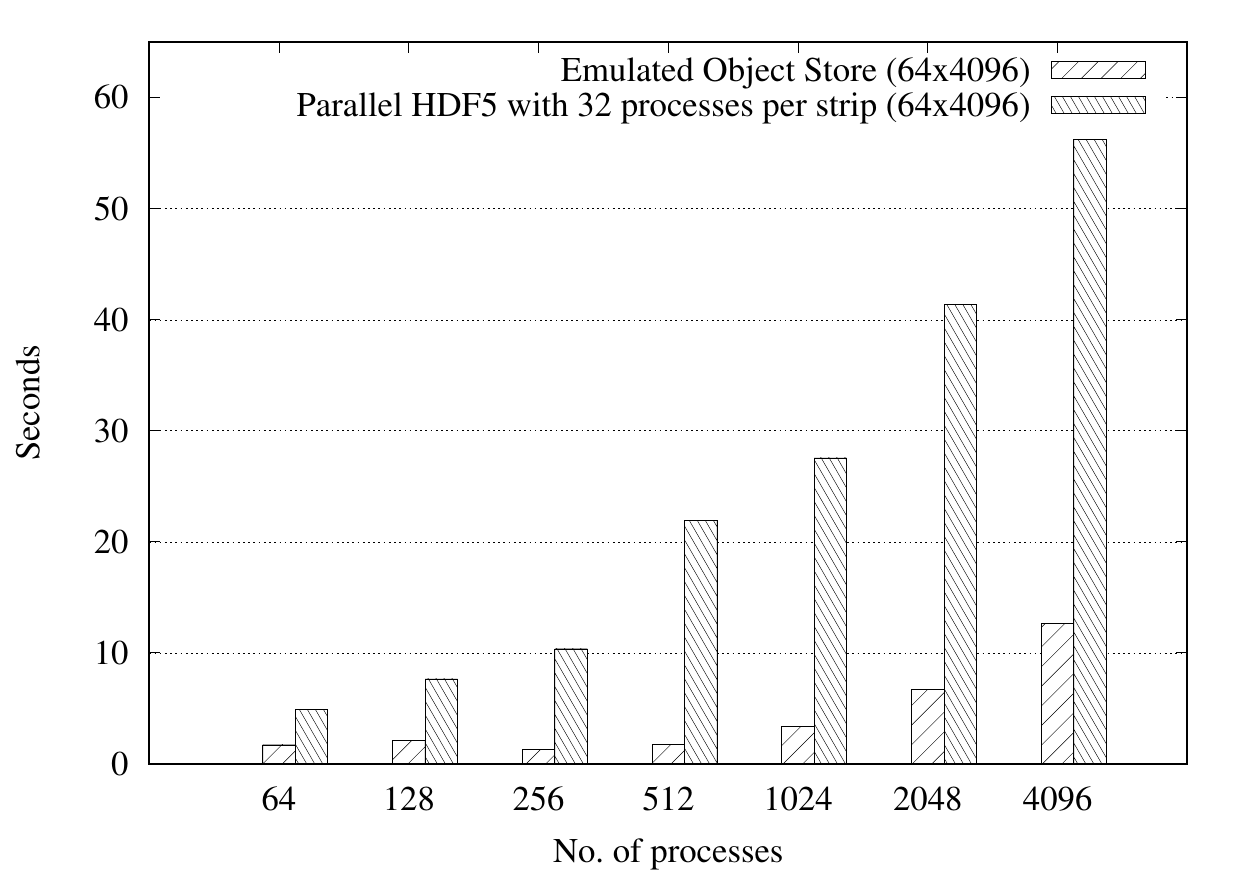}
		\caption{Average total time spent on I/O in seconds with chunk size $64 \times 4096$. For parallel HDF5 32 processes per strip is set.}
		\label{fig:io-time-figure}
	\end{center} 
\end{figure}
Fig.~\ref{fig:io-time-figure} shows the total time spent on I/O during application execution. We observe a large increase of time spent on I/O relative to number of processes used for the implementation with parallel HDF5. On the other hand, the implementation with our emulated object store shows relatively little change in terms of time spent.

%% file: conclusion.tex
\section{Conclusion}
\label{sec:conclusions} 
One of the performance and scalability bottlenecks in large scientific applications is parallel I/O to file systems. In fact, the usage of parallel file systems and consistency requirements of POSIX (that all the traditional HPC parallel I/O interfaces adhere to) poses limitations to scientific applications. Object storage is a promising technology that could address the parallel I/O scalability issues at extreme scale. In this work, we designed and implemented a library to emulate the object storage operation semantics with the goal of understanding the scalability benefit of scientific HPC applications, using object storage on large scale supercomputers. We showed that scientific applications can benefit from the usage of object storage on large scales.

In the future, we would like to apply our library to HPC applications with heavy I/O workload patterns and investigate further for potential improvements. In particular, we would like to implement a submodule with IOR~\cite{shan2008characterizing}, an I/O benchmarking software such that we can perform qualitative studies of how object store I/O semantics can contribute to scalability. Through these studies, we also hope to identify the requirements for HPC oriented object stores and how they can contribute to future highly parallel systems.